# Non-invasive transmission electron microscopy of vacancy defects in graphene produced by ion irradiation

Ossi Lehtinen[1], I-Ling Tsai[2], Rashid Jalil[2], Rahul R. Nair[2], Juhani Keinonen[3], Ute Kaiser[1*], Irina V. Grigorieva[2*]

[1]Central Facility for Electron Microscopy, Group of Electron Microscopy of Materials Science, University of Ulm, 89081 Ulm, Germany

[2]Manchester Centre for Mesoscience & Nanotechnology, Manchester M13 9PL, UK

[3]Department of Physics, P.O. Box 43, FI-00014 University of Helsinki, Finland

**Irradiation with high-energy ions has been widely suggested as a tool to engineer properties of graphene. Experiments show that it indeed has a strong effect on graphene's transport, magnetic and mechanical characteristics. However, to use ion irradiation as an engineering tool requires understanding of the type and detailed characteristics of the produced defects which is still lacking, as the use of high-resolution transmission microscopy (HRTEM) – the only technique allowing direct imaging of atomic-scale defects – often modifies or even creates defects during imaging, thus making it impossible to determine the intrinsic atomic structure. Here we show that encapsulating the studied graphene sample between two other (protective) graphene sheets allows non-invasive HRTEM imaging and reliable identification of atomic-scale defects. Using this simple technique, we demonstrate that proton irradiation of graphene produces reconstructed monovacancies, which explains the profound effect that such defects have on graphene's magnetic and transport properties. This finding resolves the existing uncertainty with regard to the effect of ion irradiation on the electronic structure of graphene.**

Knowing the detailed microscopic structure of atomic-scale defects in graphene, such as vacancies or grain boundaries, is crucially important for understanding and potentially controlling their effect on electronic and spin transport, mechanical strength, chemical reactivity and thermal conductivity of this versatile material[1-8]. For example, vacancies produced by high-energy ion irradiation have been shown to drastically reduce electron mobility, leading to poor conductivity of graphene devices[3]. On the other hand, magnetic moments associated with monovacancies[8-10] or voids[11] in graphene make it potentially useful for spintronic devices, either as a way to control spin currents[4,10] or as a magnetic component, if a reliable method of inducing ferromagnetic coupling between these local moments is found. Furthermore, it was suggested that grain boundaries can be used to engineer local transport gaps in graphene[12] while line defects can behave as metallic one-dimensional wires[13] or act as filters for charge carriers according to which valley they occupy, thereby creating an opportunity to develop unconventional electronic applications, so-called valleytronics[14].

The possibility to use atomic-scale defects in order to engineer graphene properties is underpinned by defect-induced changes in its band structure, for example, the appearance of sharp peaks in the density of states (localized states) near the Dirac point in the case of vacancies[1,11] or changes in electron/hole scattering efficiency at grain boundaries and line defects[12-14]. In turn, the changes in the electronic structure strongly depend on detailed characteristics of the defects, e.g. the presence of dangling bonds, saturation of dangling bonds, possible reconstruction in the case of vacancies, and periodicity and specific type of defects making up a grain boundary. Irradiation with high-energy ions is one of the most widely used methods to introduce defects in graphene (e.g. refs.[3,6,9,15]) yet the exact atomic structure of irradiation-induced vacancies – and, accordingly, their effect on graphene's electronic structure – remains unknown and a matter of debate. For example, recent observations of magnetic behavior of proton-irradiated graphene[8-10] can only be understood if the majority of irradiation defects are reconstructed monovacancies (i.e. having undergone Jahn-Teller distortion[16]), rather than e.g. divacancies, Stone-Wales- or other complex defects. Yet there is no experimental evidence to support this assumption[17-19]. On the contrary, it is often assumed that, due to a high reactivity of dangling bonds, monovacancies coalesce and transform into divacancies or more complex 555-777 defects[20,21] or the dangling bonds become saturated with e.g. hydrogen preventing the Jahn-Teller reconstruction[16,22].

Modern high-resolution transmission electron microscopy (HRTEM) offers a technique capable of visualizing individual (even light) atoms, and a single missing atom can be detected even at relatively low accelerating voltages ≤80 keV[23-27]. This is due to recent advances in hardware aberration correction[28,29] and in minimizing the spread of the atom contrast in the image[30-33]. However, very high electron doses are required to achieve a signal-to-noise ratio sufficient for atomic resolution[34]. Under such conditions, sputtering of individual carbon atoms has been observed as well as frequent transformations of defect structures, e.g., through the bond-rotation mechanism[35,36]. As a result, the observed defects are either modified or even created by the electron beam used to observe them, rather than being intrinsic[35-41]. Monovacancies are particularly unstable under the continuous electron bombardment and quickly transform into, *e.g.*, divacancies through knocking out of the single undercoordinated carbon atom[20,38,40]. In addition, contaminants on the sample surface and residual gases in the vacuum of the microscope can be broken down, creating free radicals and leading to chemical reactions, especially in the vicinity of defect sites that have a high affinity to contamination[42]. Observations show that graphene samples with high defect densities tend to get covered in contamination when exposed to ambient conditions, which further inhibits characterization of the defects. Taken together, these factors have made it impossible so far to reliably characterize defects in graphene, whether intrinsic or created by irradiation.

In this contribution we use aberration-corrected transmission electron microscopy (AC-HRTEM) to show that encapsulation of defective graphene between two other graphene sheets overcomes the above difficulties and allows reliable identification of vacancy-type defects created by proton irradiation. The outer graphene layers isolate the studied defects from external species that would otherwise react with the dangling bonds, inhibit sputtering of carbon atoms, and provide an ideal protective coating against radiation damage due to the exceptionally high electrical and thermal conductivity of graphene, its chemical stability, transparency and crystallinity[43,44]. Recently graphene encapsulation has been used successfully in TEM to observe growth of nanocrystals[43] and study soft-hard interfaces[44], as well as to study radiation-sensitive monolayers of $MoS_2$ sandwiched between two layers of graphene[45,46]. Separating the contrast originating from protective graphene layers and from the encapsulated material in refs.[43-46] was easy, due to the large difference in atomic numbers for the studied- and the protective layers. For all-graphene sandwiches, as in the present study, this is nontrivial. In the case of turbostratic (non-aligned) stacking of the layers, the ideal honeycomb lattice of the outer layers can be removed from the micrographs by Fourier filtering in digital post-processing, similar to refs.[43-45], but identification and positioning of the defects is not always possible. In the case of perfect ABA stacking of the layers, the Fourier filtering approach is not feasible. Nevertheless, as we show below, such stacking presents an ideal case for atomically resolving defects in the encapsulated graphene, by working at optimized defocus conditions.

Using this new approach we demonstrate that proton irradiation produces monovacancies with simple Jahn-Teller reconstruction, which is the first experimental evidence in support of the proposed explanations for the effect of ion irradiation on transport and magnetic properties of graphene[3,4,8-10]. The type of ions (protons) and energy range (350 keV) used in our experiments are typical for many studies of defect-dependent electronic, magnetic and other properties of graphene. Therefore, in addition to demonstrating a non-invasive method to study graphene's atomic structure, our findings impact significantly on the general understanding of the behavior of graphene devices under irradiation.

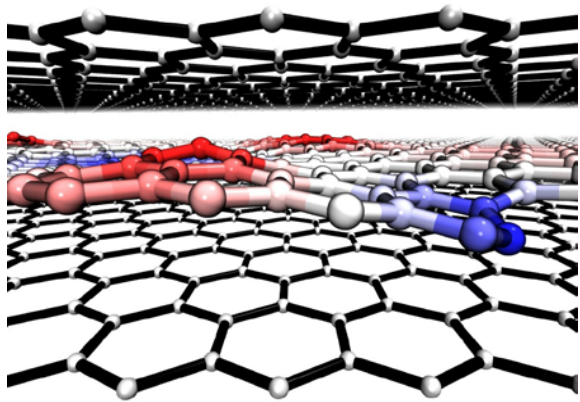

**Figure 1.** Schematic representation of the sandwich structure (defected graphene between two protective graphene layers). Colored areas of the middle layer represent defects.

**Results and discussion.**

The principle of encapsulating defective graphene between protective outer layers is illustrated in Fig. 1 where colored areas indicate defect sites. To fabricate such samples for HRTEM study, graphene flakes were first mechanically exfoliated onto $Si/SiO_2$ and mono-, bi- and trilayer flakes identified using optical microscopy[47] and Raman spectroscopy (corresponding Raman spectra are given in the Supplementary Information). The flakes were then transferred onto TEM grids following the previously developed procedure[48,49] (Supplementary Figure S1), also described in Methods. After that the samples were irradiated with 350 keV protons using a 500 kV ion implanter system (see Methods for details). The energy, fluence and other irradiation parameters were chosen to be as similar as possible to those used in earlier experiments on vacancy induced magnetism in graphene[9,10], to ensure that direct comparisons can be made with the results of those studies. The HRTEM micrographs were recorded using hardware aberration-corrected FEI TITAN 80-300 operated at 80 kV. The typical pre-TEM heat treatment of the samples was omitted in these experiments in order to avoid migration, coalescence or recombination of point defects in the samples.

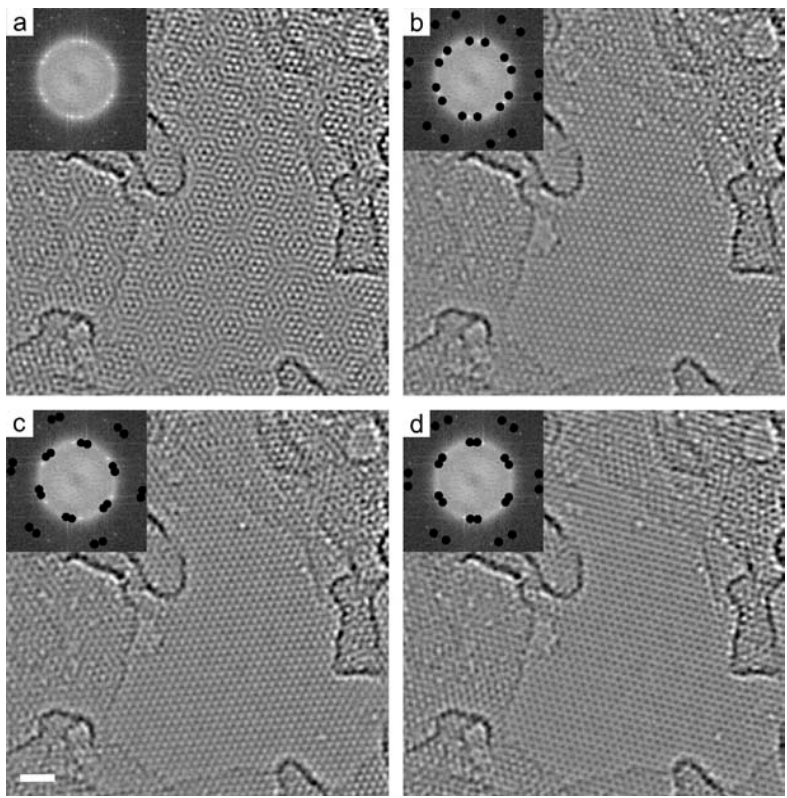

**Figure 2. Graphene oxide sandwiched between two protective graphene monolayers under the electron microscope.** (a) shows the acquired 80kV AC-HRTEM image displaying a Moire pattern characteristic of three overlapping, misoriented layers of graphene as indicated by the FFT in the inset which shows three sets of diffraction spots from graphene's hexagonal lattice. In (b-d) regular honeycomb lattices of the two layers corresponding to two sets of diffraction spots indicated by black dots (applied masks) have been digitally removed by Fourier filtering. Note the inverted contrast of the graphene lattice in (d), a signature of the bilayer. The scale bar corresponds to 1 nm.

To prove the possibility of identifying defects in multilayer samples schematically shown in Fig. 1, we first prepared a bilayer of graphene oxide (mechanically exfoliated from graphite oxide), known to contain significant amounts of atomic-scale defects, which was sandwiched between two encapsulating graphene monolayers (Methods). This resulted in a turbostratic graphene heterostructure. A corresponding HRTEM image in Fig 2a shows a Moire pattern typical of overlapping misoriented layers of graphene. To extract images of individual layers, we applied digital post-processing Fourier filtering where two perfect graphene lattices, corresponding to two of the three sets of diffraction spots, are removed in each image of Fig. 2b-d as indicated by black spots masking the corresponding reflections. This revealed – in addition to the ~20nm clean area showing atomically resolved graphene lattice – several point defects (bottom right corner and near the top of the clean area). Importantly, neither the position, nor the appearance of the defects changed during imaging, indicating the absence of atomic displacements under the electron beam. Nevertheless, this experiment showed that simply imaging a whole sandwich under standard conditions (Scherzer focus) and applying post-processing does not allow reliable identification of either the type or the location of defects: Indeed, the defects in Fig. 2 are likely to be located in only one of the layers, as their positions are the same in Fig. 2b, c and d, but they still appear in all three images, giving slightly different contrasts. This is because the defects contain spatial frequencies different from those for the perfect honeycomb lattice and artificially removing a regular continuous lattice for a layer that contains a point defect will leave an 'imprint' of the defect in other layers, even if the latter are defect free.

Below we show that this limitation can be overcome by using protective graphene layers that are in perfect stacking with the studied middle layer (i.e. using ABA stacked trilayer graphene) and by imaging at an optimized defocus (~ -15 nm in our case, the exact value depending on the spherical aberration coefficient of the microscope). Combined with corresponding image simulations, the obtained HRTEM images allow unambiguous identification of the defects. To this end, Fig. 3 shows a simulated focal series for three different point defects residing in the middle layer of a trilayer graphene sample (see Methods for details of the simulation technique). At Scherzer focus (the leftmost column) only faint changes in contrast can be observed at defect sites because the signal from the periodic structure of the other layers obscures the defects. However, as the focus is adjusted away from the 'optimal' conditions, the signal arising from the periodic graphene structure is suppressed, and the less regular defect sites become more pronounced – see Fig. 3, and also a larger collection of defects in Supplementary Fig. S4. The differences between different types of defects (reconstructed and non-reconstructed monovacancies, divacancies, 555-777 and 585 defects, Stone-Wales defects, adatoms) become very clear at larger defocus and it is also clear that the defocus can be optimized to maximize these (compare simulated images for different types of defects in Figs. 3 and S4 at the optimum defocus 14.5-15 nm). We emphasize that the above simulated images correspond to a trilayer graphene crystal with a perfect ABA stacking between the two outer protective layers and the middle layer. This represents an extreme case where the digital removal of the images of two outer layers (as in Fig. 2) is no longer possible yet the defects can be clearly identified and fully characterized. Below we use this technique to identify point defects created by proton irradiation in graphene samples.

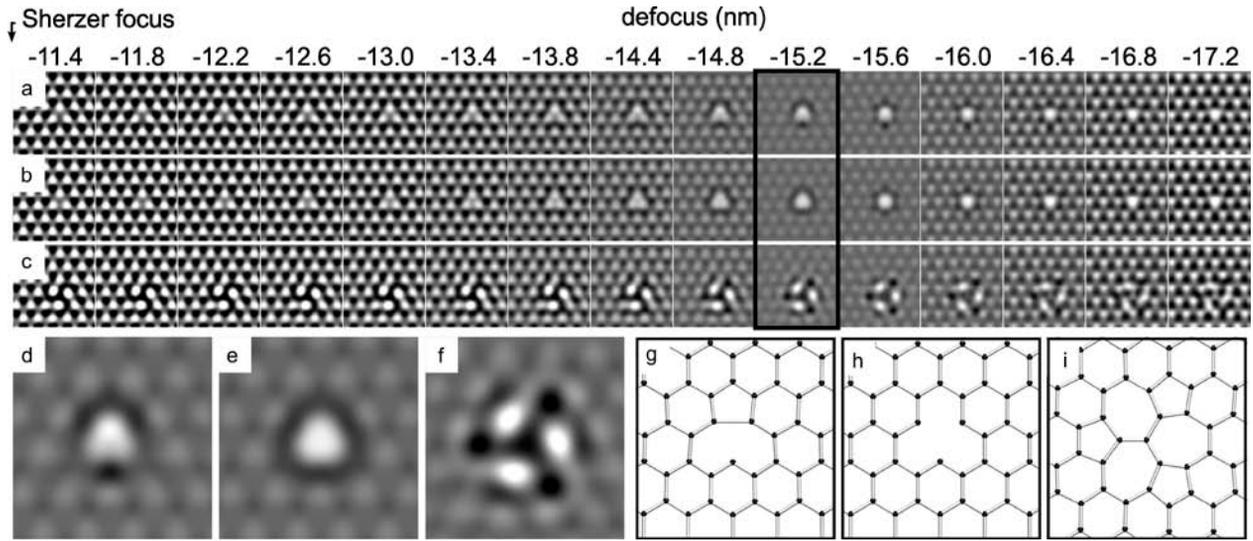

**Figure 3**: **Revealing graphene defects.** Simulated 80kV AC-HRTEM images of point defects in trilayer graphene at different focusing conditions: (a) reconstructed and (b) unreconstructed monovacancy; (c) divacancy (555-777 defect). The structural models were relaxed using the DFTB formalism and taking into account the spherical aberration. Each row shows a focal series for the same defect, at a progressively increasing defocus. (d-f) magnified images of the reconstructed and unreconstructed monovacancy and a divacancy at the optimum defocus (-15.2 nm). (g-i) schematic representation of the arrangement of atoms for the reconstructed (g) and unreconstructed (h) monovacancy and a divacancy (i).

Figures 4 and 5 show several examples of defects found in trilayer samples. We note that the defect separation averaged over all acquired images was 8.4 nm, which explains that just one or two defects are present in each of the ~10×14 nm panels in Figs. 4 and 5. The observed defect separation is in excellent agreement with the expected defect density of 8.1 nm for 350 keV proton irradiation and a fluence of $2\times10^{16}$ ions·cm$^{-2}$ as used in our experiments. The latter was derived by calculating the displacement cross-section for the graphene-proton interaction using the ZBL-repulsive potential and assuming a displacement threshold of 28 eV (the standard value for graphite used in SRIM software package[50]).

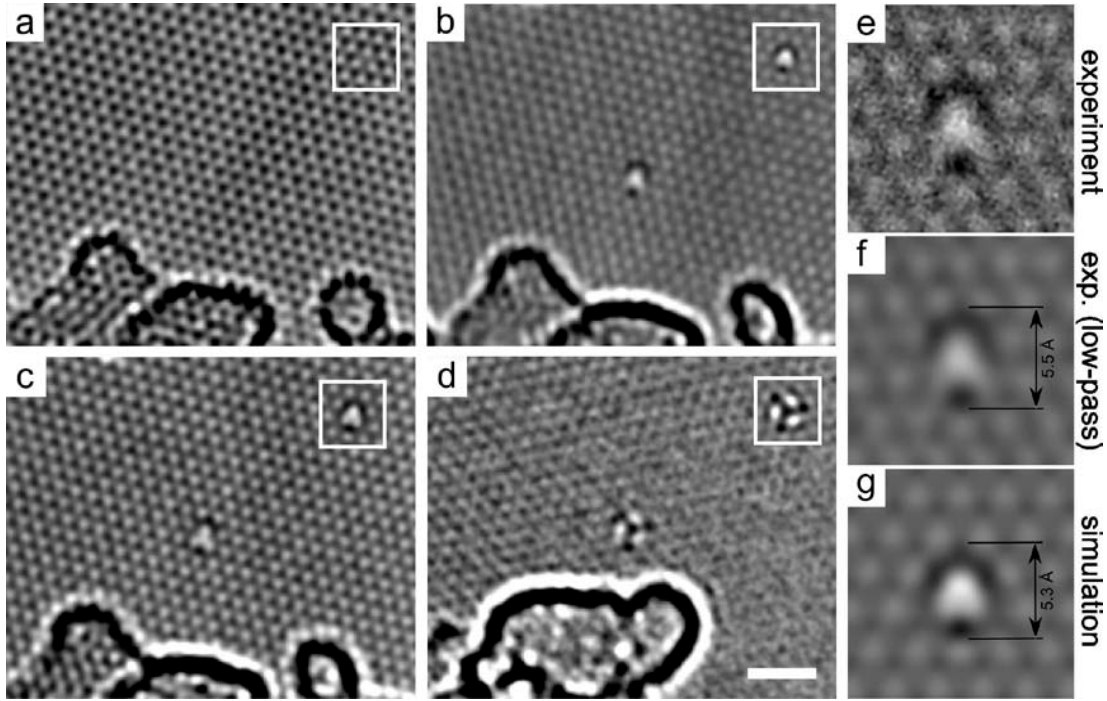

**Figure 4**. **Reconstructed monovacancy: optimizing the view.** (a-c) 80kV AC-HRTEM images of a reconstructed monovacancy in the middle layer of proton-irradiated triple layer graphene at three different focusing conditions: (a) Scherzer focus, (b) approximately 15 nm and (c) 16 nm underfocus. (d) AC-HRTEM image of a divacancy (555-777 defect). Insets show corresponding simulated images. (e-g) Detailed analysis of an AC-HRTEM image of a reconstructed monovacancy: (a) raw image obtained with 15 nm underfocus; (f) the same image after application of the low-pass filter; (g) corresponding simulated image for 14.8 nm underfocus. The scale bar corresponds to 1 nm.

Fig. 4a-c shows an example of a reconstructed monovacancy (the corresponding arrangement of atoms - two initial dangling bonds are saturated in the reconstruction, creating a highly asymmetric defect[51] – is shown schematically in Fig. 3g). The three experimental images show the same area of the sample recorded at different focusing conditions, going from the Scherzer focus in panel (a) to approximately -15 and -16 nm defocus in panels (b) and (c), respectively. The strong asymmetry characteristic for a reconstructed monovacancy is most obvious at -15nm defocus. At a defocus approximately 1 nm away from the optimal value [panel (c)] the distinctive features of the reconstructed monovacancy are less pronounced but still present. At both defocusing conditions the experimental images match very well the corresponding simulations for a reconstructed monovacancy (shown as insets) thus allowing unambiguous identification of the defect. To emphasize this point, we compared simulated and experimental images of the monovacancy in greater detail, as shown in panels (e-g) of the same figure, which produced excellent quantitative agreement between the experiment and simulations.

As clear from simulated images, each type of vacancy has a set of unique features which we used to identify different defects in our samples. For example, the image of a reconstructed monovacancy is elongated, has short dark 'wings' at one end and an elongated bright spot at the centre. In contrast, the image of an unreconstructed monovacancy is a symmetric, almost triangular, dark contour with a round bright spot in the middle. Fig. 4d shows an image of a 555-777 (unreconstructed) divacancy, which is very distinctive and clearly different from the monovacancies. The image of a reconstructed divacancy (585 defect, shown in Supplementary Figure S4) is asymmetric, somewhat similar to the reconstructed monovacancy, but still easy to distinguish from the latter, as it is 50% more elongated and has long dark 'wings' next to a round bright spots in the centre.

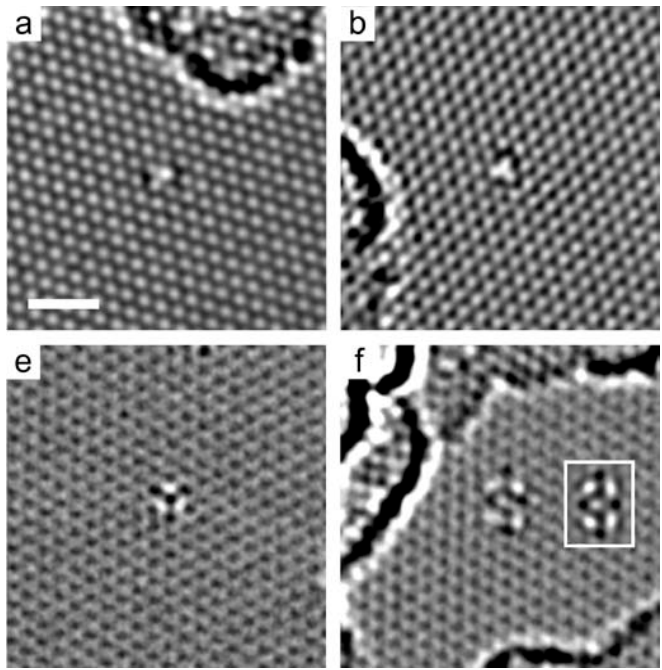

**Figure 5**. **Further examples of defects produced by proton irradiation.** (a,b) reconstructed monovacancies; (e) a 555-777 divacancy; (f) a 5555-6-7777 defect (inset shows a corresponding simulated image). The scale bar corresponds to 1 nm.

Analysis of all found defects showed that a significant[52] proportion were monovacancies (always reconstructed), with only a few divacancies found in all studied samples. This is in contrast to earlier HRTEM observations on unprotected monolayer graphene (e.g. refs.[21,37-39]) where the majority of point defects were identified as either divacancies or more complex defects, which were also unstable under the electron irradiation during imaging. We believe this is a direct manifestation of the protective effect of the two outer graphene layers that prevent transformation of monovacancies into other types of defects.

Fig. 5 shows further examples of point defects found in other areas. The focusing conditions here are close to those in Fig. 4c and the defects in panels (a,b) have been identified as reconstructed monovacancies, panel (e) shows a 555-777 divacancy and panel (f) a 5555-6-7777 defect (cf. Supplementary Fig. S4).

We emphasize that in all the above observations the defects were stable under the electron beam, at least for a typical time required to record an image sequence, several minutes. At the same time, no stable monovacancies could be found in mono- or bilayer graphene prepared and irradiated together with the trilayers, in agreement with other HRTEM studies. Furthermore, all the observed defects were the result of irradiation, not e-beam damage: Indeed, same type of defects were found in several samples irradiated under identical conditions but at different times (time between irradiation and observation differing by several weeks). At the same time, none of these defects were found in non-irradiated samples. We can therefore conclude that using trilayer graphene instead of the usual monolayers provides, for the first time, an opportunity to study truly intrinsic defects in graphene.

Finally, let us discuss another important issue in atomic-resolution TEM studies of graphene, the issue of contamination. It is well known that as-exfoliated graphene is always covered in a layer of contamination (hydrocarbons) with only small atomically clean areas in the range of 10-20 nm, where atomic resolution imaging can be conducted. The contamination problem becomes much worse when conducting ion irradiation treatments due to the often sub-optimal vacuum conditions, and sample surfaces tend to be completely covered in contamination after irradiation. This was also the case in the present study: Despite the extra care taken during preparation of the first batch of samples for irradiation (see Methods), already the first TEM observations showed that the layer of contamination was too extensive to achieve atomic resolution in desired parts of the samples, as only a few clean ~5nm spots could be located in an entire ~100μm size sample. *In situ* annealing of the samples is typically used for removing hydrocarbon buildup prior to imaging, but it could not be applied in our case, as this would modify the defect structures. However, we found that contamination could be drastically reduced by depositing a small amount of nanoscale crystallites of two-dimensional $MoS_2$ onto the graphene samples prior to proton irradiation (see Methods for details). The effect of $MoS_2$ presence is clear from comparing the two low-magnification images in Fig. 6: While clean areas in $MoS_2$-free sample are so small (<2 nm) as to be almost invisible on this scale, samples with $MoS_2$ flakes deposited on top are mostly clean graphene with uninterrupted clean patches >20nm in size. This was the case for all graphene flakes (mono-, bi- and trilayers) and made locating point defects relatively easy.

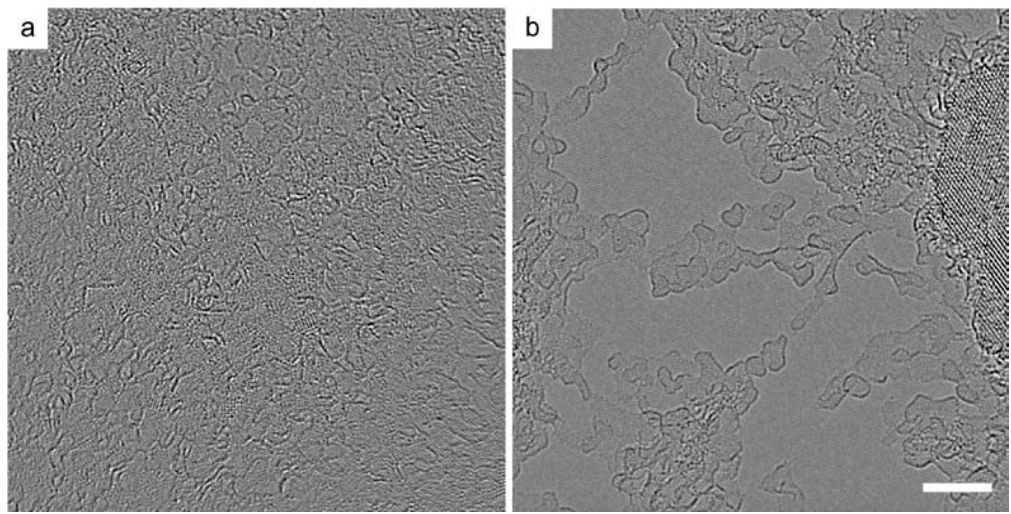

**Figure 6**. **Cleaning effect of $MoS_2$.** Overview AC-HRTEM images of irradiated triple layer graphene with and without $MoS_2$ crystallites deposited on top of graphene before irradiation (right and left panels, respectively). Only small clean patches are visible in panel (a), whereas the clean areas are much larger in panel (b) [a few overlapping crystalline MoS2 flakes are visible on the far right of (b)]. The scale bar corresponds to 5 nm.

We emphasize that the presence of randomly placed $MoS_2$ flakes was the only difference between the two batches of samples – see Fig. 6 – and they were all irradiated in the same run. The images of monovacancies in Figs. 4 and 5 were obtained on the trilayer flake shown in the right panel of Fig. 6. The exact mechanism behind this curious cleaning effect of $MoS_2$ is not known but we speculate that it is likely to be due to the known catalytic activity of $MoS_2$ with respect to hydrogen evolution[53,54] which helps cracking the hydrocarbons on the graphene surface without the need of annealing. $MoS_2$

flakes in our case were obtained by liquid exfoliation in an organic solvent which produced a suspension of 10-20nm flakes that were then deposited onto graphene by drop-casting. The amount of $MoS_2$ flakes can be easily controlled by varying their concentration in the suspension as described in Methods. This simple technique provides a valuable alternative to the standard removal of contamination from graphene by annealing and can be particularly useful in studies of intrinsic defect structures where annealing is not possible.

In conclusion, using the example of vacancies produced by proton irradiation, we show that encapsulating the studied graphene sample between two other (protective) graphene sheets allows non-invasive HRTEM imaging of atomic-scale defects and their reliable identification. We demonstrate that this is possible even in the extreme case of perfect ABA stacking of the three graphene layers, i.e. using trilayer - instead of the usual monolayer - graphene. The defects in the protected middle layer can be reliably identified by imaging under optimized defocus conditions (typically 14-15 nm) and analyzing the HRTEM micrographs in conjunction with image simulations. While practically indiscernible under standard focusing conditions (Scherzer focus), defects become visible with larger defocus as it effectively 'filters out' the regular honeycomb lattice revealing the imperfections. Using this technique, we demonstrate that reconstructed monovacancies are produced in significant numbers by proton irradiation under typical conditions, which explains the profound effect that such defects have on graphene's magnetic and transport properties. This finding resolves the existing uncertainty with regard to the effect of ion irradiation on the electronic structure of graphene.

**Methods**

Graphene flakes ~100 μm in size were first mechanically exfoliated from natural graphite onto an oxidized $SiO_2$/Si substrate (from IDB Technologies Ltd: 290 nm $SiO_2$, n-type doped, one side polished). The monolayer and trilayer flakes were located and identified by optical microscopy as shown in Supplementary Figure S1a. For transfer from $SiO_2$/Si substrate to the TEM grid, a layer of e-beam resist PMMA (polymethylmethacrylate) (MicroChem, 950,000 MW, 3 wt. % in anisole) was spin coated onto the substrate. The PMMA layer with all graphene fakes attached to it was then detached from the substrate by partially etching the underlying $SiO_2$ surface with an aqueous solution of KOH (0.5M). After several rinsings with deionised water, the PMMA-graphene membrane was transferred onto a TEM grid. Finally the PMMA layer was dissolved in acetone and the TEM grid with graphene flakes was dried in a critical point dryer (CPD) – see Supplementary Figure S1b for an image of the resulting sample.

To fabricate samples covered with a small amount of dispersed $MoS_2$ crystallites (see Supplementary Information), we first mechanically exfoliated graphene onto oxidized silicon as above [Supplementary Figure S2(a)]. To prepare thin (few-layer) $MoS_2$ we used a well known technique of liquid exfoliation[55] to obtain 0.1gm/L dispersion of $MoS_2$ in N-methyl pyrrolidone (NMP). To this end, 50 mg of $MoS_2$ powder (Sigma Aldrich) was sonicated in 60 ml of NMP for 20h. The obtained suspension was centrifuged to remove large multilayer crystallites and then carefully dropcast over the graphene flakes on $SiO_2/Si$. After that the substrate was heated at 70 °C for 10 min and then cleaned using standard solvent treatment (acetone and isopropyl alcohol (IPA)) – see Supplementary Figure S2(b). Finally graphene covered with the remaining $MoS_2$ crystallites were transferred onto a TEM grid following the procedure described above – See Supplementary Figure S2(c).

The samples were irradiated using a 500 kV ion implanter using 350 keV protons at a total ion fluence of $2 \cdot 10^{16}$ ions/cm$^2$. To achieve uniform irradiation of all graphene flakes on a TEM grid, the accelerated proton beam was rasterized over the sample area. All irradiations were done at room temperature and current densities <0.2 µA/cm$^2$.

Aberration-corrected high-resolution (AC-HR)TEM imaging was carried out in an FEI Titan 80–300 transmission electron microscope equipped with an objective-side image corrector. The microscope was operated at 80 kV. The extraction voltage of the field emission source was set to a reduced value of 2 kV in order to minimize the energy spread of the electron beam. The spherical aberration was set to 20 µm. Most of the images were taken with strong underfocus to suppress the visibility of the graphene lattices and enhance the visibility of the defects. The resulting image sequences were background subtracted and drift compensated. All images except Fig. 6 (single-frame images) were averaged over 3–10 frames (in which the atomic structure did not change) to improve the signal-to-noise ratio.

The structural models for the image simulations were relaxed using the density functional tight binding formalism[56,57] (the interlayer Van der Waals interactions were not taken into account). The relaxations were conducted via a damped molecular dynamics simulation. Single k-point was used in the calculations at 0 K temperature . The relaxed structures were used for HRTEM image simulations where we employed the QSTEM software[58]. The image simulations were conducted for 80 keV electrons, with spherical aberration coefficient of 0.02 mm and a focal spread of 6 nm.

**Acknowledgements.** This work was supported by the UK Engineering and Physical Sciences Research Council, the DFG (German Research Foundation) and the Ministry of Science, Research and the Arts (MWK) of Baden-Wuerttemberg in the framework of SALVE (Sub Angstrom Low-Voltage Electron Microscopy) project. O.L. acknowledges support from the Finnish Cultural Foundation.

**Supplementary Information Available:** Four supplementary figures with captions.

# Supplementary Information

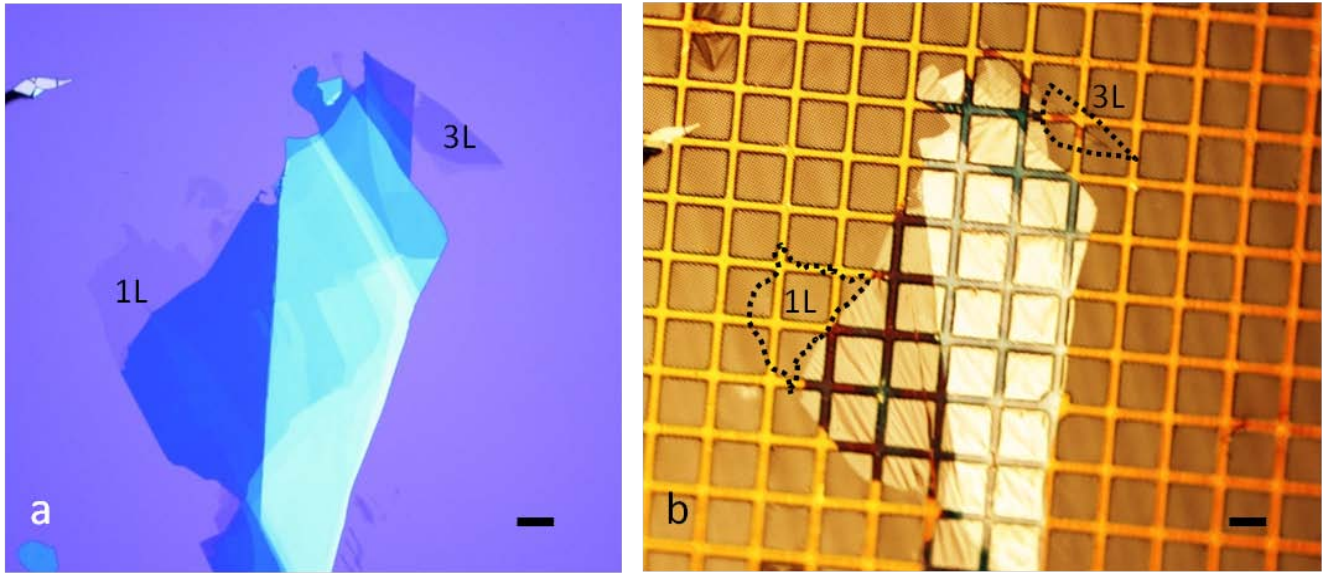

**Supplementary Figure S1**: **Isolating mono- and trilayer graphene.** (b) Optical image of a large graphene flake on Si/ $SiO_2$ substrate with a monolayer region (marked 1L) and a trilayer region (marked 3L)  (b) Optical image of the same flake after transfer onto a TEM grid. The mono- and trilayer regions are shown with doted black lines.  Scale bar corresponds to 50μm on both images.

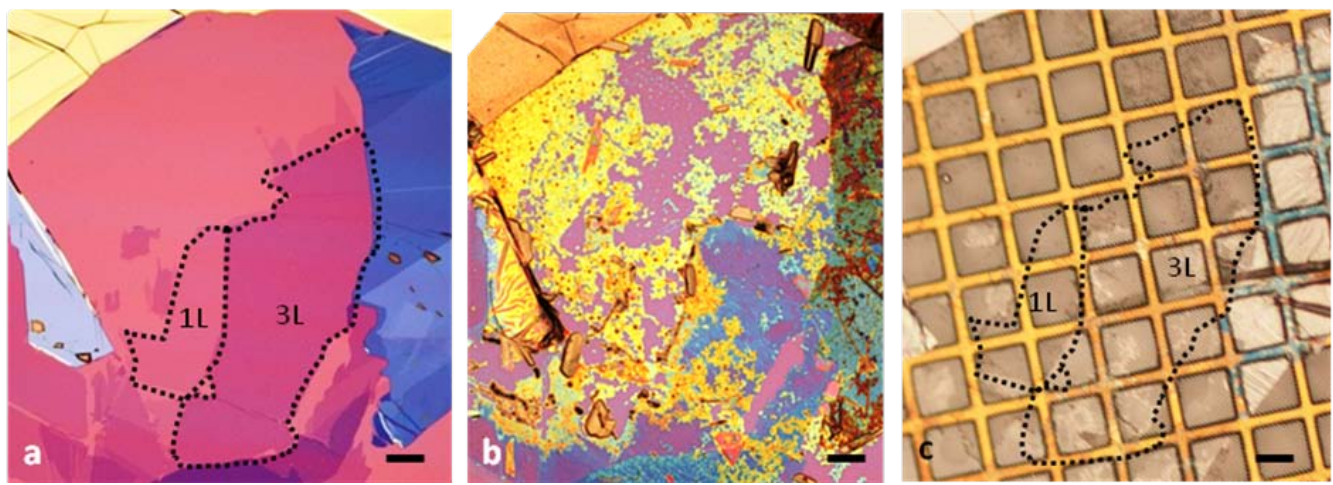

**Supplementary Figure S2.** (a) Optical image of graphene flakes on $SiO_2/Si$ substrate with a monolayer region marked 1L and a large trilayer region marked 3L. (b) $MoS_2$ flakes deposited on top of graphene; blue and yellow colours indicate thin and thick layers of $MoS_2$ flakes, respectively. (c) optical image of $MoS_2$-graphene flakes after transfer onto the TEM grid. Scale bars correspond to 50μm.

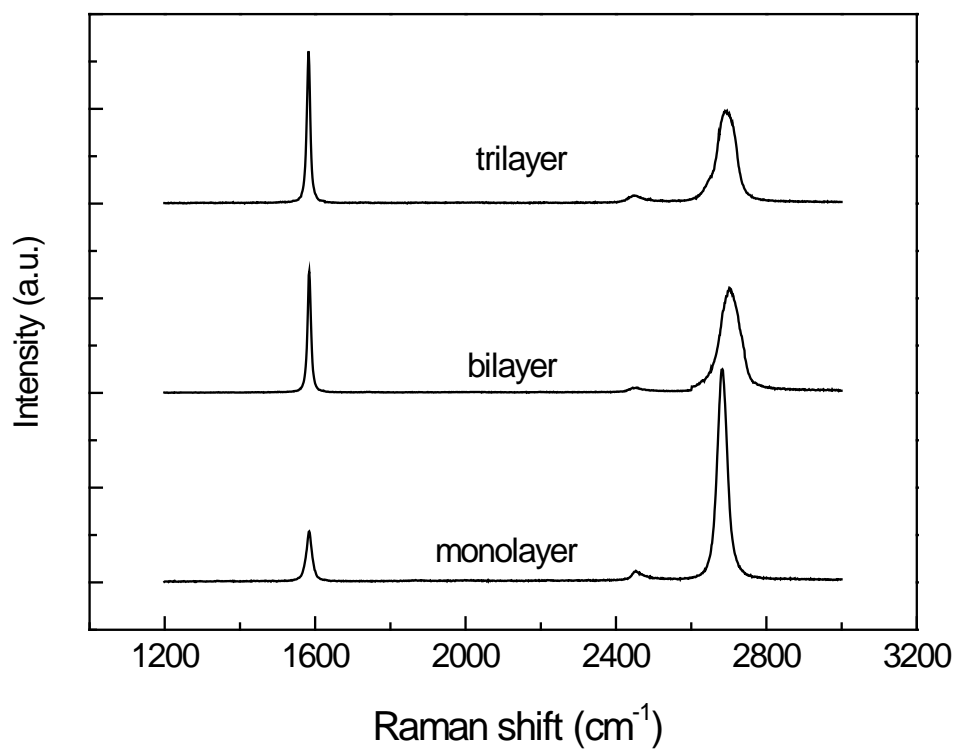

**Supplementary Figure S3. Identification of the number of layers in studied graphene samples.**

The number of layers in each sample was determined using Raman spectrometry. The above spectra were obtained on mono-, bi- and trilayers shown in Figs. S1 and S2.

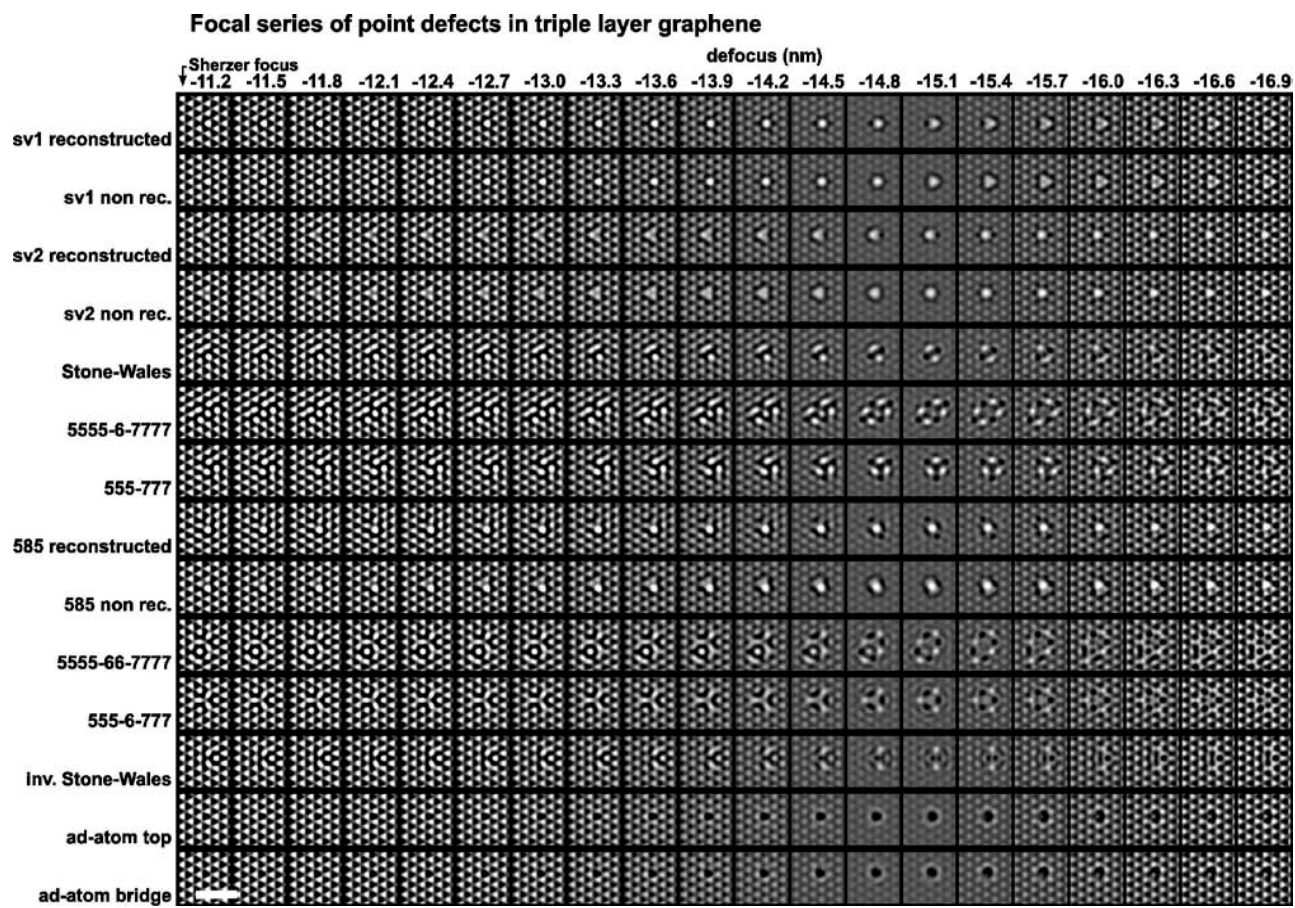

**Supplementary Figure S4.** Simulated 80kV AC-HRTEM images of a variety of defects in triple layer graphene at different focusing conditions and taking into account spherical aberration. The structural models were relaxed using the DFTB formalism. Each row shows a focal series for the same defect, at a progressively increasing defocus. The images correspond to defects residing in the middle layer of a trilayer graphene sample. 'sv1' and 'sv2' refer to single vacancies where a carbon atom was removed from sublattice A and B, respectively. The scale bar corresponds to 1 nm.